\documentclass[conference]{IEEEtran}
\IEEEoverridecommandlockouts
\usepackage{cite}
\usepackage{amsmath,amssymb,amsfonts}
\usepackage{algorithmic}
\usepackage{graphicx}
\usepackage{textcomp}
\usepackage{xcolor}
\def\BibTeX{{\rm B\kern-.05em{\sc i\kern-.025em b}\kern-.08em
    T\kern-.1667em\lower.7ex\hbox{E}\kern-.125emX}}

\usepackage{amsthm}

\newtheorem{theorem}{Theorem}[section]
    
\newtheorem{lemma}[theorem]{Lemma}
\theoremstyle{remark}
\newtheorem*{rem}{Remark}
    
\begin{document}

\title{Ordered Transmission-based Detection in Distributed Networks in the Presence of Byzantines }

\author{\IEEEauthorblockN{Chen Quan$^{\dagger}$,Saikiran Bulusu$^{\dagger}$, Baocheng Geng$^{\ddagger}$, and Pramod K. Varshney$^{\dagger}$}\\
	\IEEEauthorblockA{$^{\dagger}$ Department of EECS, Syracuse University, NY, 13244 USA  \\
		$^{\ddagger}$ Department of Computer Science, University of Alabama at Birmingham, AL, 35294 USA\\
		Email: chquan@syr.edu, sabulusu@syr.edu,  bgeng@uab.edu,  varshney@syr.edu }\\
}

\maketitle

\begin{abstract}
The ordered transmission (OT) scheme reduces the number of transmissions needed in the network to make the final decision, while it maintains the same probability of error as the system without using OT scheme. 
% In practice, some security issues might affect both the detection performance and the saving performance of the system that uses the OT scheme. 
In this paper, we investigate the performance of the system using OT scheme in the presence of Byzantine attacks for binary hypothesis testing problem. We analyze the probability of error for the system under attack and evaluate the number of transmissions saved using Monte Carlo method. We also derive the bounds for the number of transmissions saved in the system under attack. The optimal attacking strategy for the OT-based system is investigated. Simulation results show that the Byzantine attacks have significant impact on the number of transmissions saved even when the signal strength is sufficiently large. %When adopting the optimal attacking strategy, the Byzantine nodes can maximize the error probability and significantly decrease the number of transmissions saved by utilizing the OT scheme.
\end{abstract}

\begin{IEEEkeywords}
Ordered transmissions, Byzantine attacks, wireless sensor networks.
\end{IEEEkeywords}
\section{Introduction}
%Distributed detection is a well studied topic in the literature\cite{hoballah1989distributed}. {\color{red}The sensors send their observation regarding the phenomenon of interest (PoI) to the fusion center (FC) and the FC makes a final decision about the presence or the absence of PoI depending on the sensors' observations.}
Energy-efficiency is an important aspect to consider while designing a network with prolonged lifetime \cite{feng2012survey}. Some notable schemes that improve energy efficiency by reducing the number of transmissions are censoring  \cite{rago1996censoring}, clustering \cite{bandyopadhyay2003energy}, sensor selection \cite{bajovic2011sensor}, ordered transmission (OT) \cite{blum2008energy}. 
% These schemes improve the energy-efficiency of the network by reducing the number of transmissions. 
Note that the schemes like censoring and OT, select and transmit highly informative observations to the fusion center (FC). In this paper, we consider the OT scheme in the distributed setup. In the OT scheme, all the sensors in the network are equipped with a timer and calculate their respective log-likelihood ratios (LLRs) based on the observations. When the timer counts down from $\frac{1}{|LLR|}$ to zero, the sensor transmits its respective LLR to the FC. Hence, the more informative sensors transmit relatively faster than the less informative ones. When the FC receives enough observations to make the final decision, the FC broadcasts a stop signal to stop the sensors from transmitting further. The sensors that have not yet transmitted their observations reset their timers for the next decision interval after they receive the stop signal.

The following are some of the works that consider the OT scheme for energy efficiency. 
% The OT scheme was first proposed in \cite{blum2008energy} where the number of transmissions could be reduced without degrading the performance of the system. 
In \cite{braca2011single}, the authors showed that just one observation was sufficient for the OT scheme to make the final decision with large number of sensors. In \cite{rawas2011energy}, the OT scheme was utilized for non-coherent signal detection where the LLR at each sensor can only take non-negative values, and a significant reduction in the number of transmissions was obtained. In \cite{hesham2012distributed}, a modified sequential detection scheme based on the OT scheme was proposed for spectrum sensing network. 
% The OT scheme accelerated the decision process without the degradation in the detection performance. 
In \cite{chen2020optimal}, the OT scheme was utilized in quickest change detection to reduce the number of transmissions without any impact on detection delay of the system. In \cite{sriranga2018energy}, the OT scheme was utilized in energy harvesting sensor networks to improve the energy efficiency of the sensors. In \cite{gupta2020ordered}, a correlation-aware OT scheme was proposed where spatial correlation between the sensors was considered.

Due to the distributed setup and the large deployment of low cost sensors, wireless sensor networks are vulnerable to attacks. Hence, the robustness of the network is another non-negligible aspect to consider. One typical attack is a Byzantine attack\cite{vempaty2013distributed} where a sensor may get compromised and is referred to as a Byzantine node. It can be reprogrammed and may send falsified data to the FC to degrade the detection performance of the system. However, an honest node sends genuine information to the FC. The Byzantine threat model has been extensively studied in target localization \cite{wei2017local}, collaborative spectrum sensing networks \cite{rawat2010collaborative} and ad hoc wireless networks \cite{moniz2012byzantine}. In this paper, we study the effect of Byzantine attacks on the OT-based system. 
%{\color{red}These papers need more explanation! We need more papers in the related works section! The flow is not right here... How about describing OT related mores in distributed setup, then papers related to Byzantines in distributed setup, and we consider OT + Byzantines?}

% In this work, different from the above existing works. 
Unlike the previous works, we evaluate the robustness of the OT-based system via the detection performance and the number of transmissions saved in the presence of Byzantine nodes. To the best of our knowledge, this is the first work that analyzes the effect of Byzantine attacks on the performance of such a system. The following are our major contributions: 

%  {\color{blue}(This needs work!)}
\begin{itemize}
    \item We derive the probability of error of the OT-based system under Byzantine attacks. 
    \item The number of transmissions saved in the system is evaluated numerically via the Monte Carlo approach in the presence of Byzantine nodes. We also derive an upper bound (UB) and a lower bound (LB) for the number of transmissions saved in the network.
    \item 
    % Since the error probability we obtain is an Gaussian mixture function, it is intractable to analyze the optimal attacking strategy via the error probability of the system. Hence, 
    We utilize the deflection coefficient (DC) as a surrogate for probability of error and investigate the optimal attacking strategy based on the DC of the system.
    \item The simulation results show that the optimal attacking strategy that maximizes the probability of error leads to a large reduction in the number of transmissions saved when utilizing the OT scheme.
\end{itemize}
% Due to the fact that it is not computationally-efficient enough, especially for the networks with a large number of sensors, 

The paper is organized as follows. We present our system model in Section \ref{sec:system_model}. We evaluate the performance of the system under Byzantine attacks and derive the bounds for the number of transmissions saved in the system with OT scheme in Section \ref{sec:proposed}. We present the simulation results in Section \ref{sec:simulation} and conclude in Section \ref{sec:conclusion}.

\section{System Model}
\label{sec:system_model}
In this section, we consider a binary hypothesis testing problem where hypothesis $\mathcal{H}_1$ indicates the presence of the signal and $\mathcal{H}_0$ indicates the absence of the signal. We consider a distributed network consisting of $N$ sensors and one FC. We consider the OT scheme to reduce the number of transmissions in the network. Let $y_i$ be the received observation at sensor $i\in\{1,2,\dots,N\}$. We assume that all the observations are independent and identically distributed (i.i.d) conditioned on the hypotheses. For sensor $i$, the observation $y_i$ is modeled as 
\begin{align}
\label{eq:obs_honest}
y_i = 	\begin{cases}
						n_i&\text{under $\mathcal{H}_0$}\\
						s+n_i&\text{under $\mathcal{H}_1$},    	
					\end{cases}
\end{align}

where $s$ is the signal strength each sensor uses and $n_i$ is Gaussian noise with zero mean and variance $\sigma^2$. We assume that $s$ and $n_i$ are independent. Note that $y_i$ is Gaussian with mean $s$ and $\sigma^2$ under hypothesis $\mathcal{H}_1$, and Gaussian with mean 0 and variance $\sigma^2$ under hypothesis $\mathcal{H}_0$. We next review the general OT scheme in the following.

\subsection{Network with OT Scheme}
Let $L_i$ denote the LLR for sensor $i$ given by
\begin{equation}
    L_i=\log\left(\frac{f_{Y_i}(y_i|\mathcal{H}_1)}{f_{Y_i}(y_i|\mathcal{H}_0)}\right),
\end{equation}
where $f_{Y_i}(y_i|\mathcal{H}_h)$ is the probability density function (PDF) of $y_i$ given hypothesis $\mathcal{H}_h$ for $h=\{0,1\}$. The prior probability of hypothesis $\mathcal{H}_h$ is $p(\mathcal{H}_h)=\pi_h$ for $h\in \{0,1\}$. The LLR-based optimal Bayesian hypothesis test at the FC is given by$ \sum_{i=1}^NL_i\overset{\mathcal{H}_1}{\underset{\mathcal{H}_0}{\gtrless}}\lambda=\log\left(\frac{\pi_0}{\pi_1}\right)$. Note that the sensor transmissions are ordered based on the  magnitude of their LLRs. We denote the magnitude of the ordered transmissions as $|L_{[1]}|>|L_{[2]}|>\ldots>|L_{[N]}|$. Hence, the sensor with $L_{[1]}$ transmits first, the sensor with $L_{[2]}$ transmits second, and so on.

The optimal decision rule \cite{blum2008energy} is given by
\begin{equation}
\left\{
\begin{array}{rcl}
    \sum_{i=1}^kL_{[k]}> \lambda+n_{UT}|L_{[k]}|&&\text{decide $\mathcal{H}_1$}\\
    \sum_{i=1}^kL_{[k]}< \lambda-n_{UT}|L_{[k]}|&&\text{decide $\mathcal{H}_0$},
\end{array}\right.
\end{equation}
where $n_{UT}$ is the number of sensors that have not yet transmitted at time $k$. The FC waits for the next transmission if it can not make the decision. 
The following assumption is made in \cite{blum2008energy} for the OT scheme.

{\em Assumption 1:} We assume that $Pr(L_i>0|\mathcal{H}_1)\rightarrow{1}$ and $Pr(L_i<0|\mathcal{H}_0)\rightarrow{1}$ when $s$ is sufficiently large.

\begin{rem}
Note that large $s$ is key to proving the result that the average number of transmissions saved by utilizing the OT scheme is lower bounded by $N/2$ in \cite[Theorem 2]{blum2008energy}. However, when $s$ is small or when there are Byzantine nodes in the system, Assumption 1 is no longer valid.
\end{rem}

\subsection{Byzantine Model}
Next, we discuss the case when the system is under attack.
We consider the worst case in that the Byzantine nodes know the true hypothesis and they attack based on this knowledge.
We also assume that the FC knows that there are $\alpha_0$ fraction of Byzantine nodes in the network. However, it does not know the behavioral identity of each sensor. We assume that the falsified observation $\tilde{y}_i$ for Byzantine node $i$ is given as
\begin{equation}
    {y_i}=\left\{
    \begin{array}{rcl}
        s+n_i-D&\text{if $\mathcal{H}_1$}\\
        n_i+D&\text{if $\mathcal{H}_0$},
    \end{array}\right.
\end{equation}
where $D$ is the attacking strength. Note that for an honest node the observation is $y_i$ from~\eqref{eq:obs_honest}. Hence in our setup, a sensor $i$ can be honest $(H)$ or Byzantine $(B)$. Next, we analyze the detection performance of the system with OT in the presence of Byzantine nodes.

\section{OT-based System with Byzantines}
\label{sec:proposed}
% To analyze the detection performance of the system with OT, we have the following Lemma.
In this section, we begin our analysis of the detection performance of the OT-based scheme in the presence of Byzantine nodes by first presenting the following Lemma which states that we can achieve the same detection performance without the OT scheme as that with OT scheme.

\begin{lemma}
\label{lem:det_perf_ot}
The detection performance of system with the optimal Bayesian decision rule is same as the one with the OT scheme. 
\end{lemma}
% \paragraph*{Lemma 1} 
% \paragraph*

\begin{proof}
The proof is relegated to Appendix~\ref{sec:proof_det_perf_ot}.
\end{proof}

% Therefore, we know that the probability of error of the system without ordering is the same as that of the OT-based system. 
Thus, we can evaluate the detection performance of the OT-based system by evaluating the detection performance of the system without ordering. 
For the system without ordering, we have $L_i=\frac{2y_is-s^2}{2\sigma^2}$ when sensor $i$ is honest $(i=H)$. When sensor $i$ is Byzantine $(i=B)$, the LLR is given as
\begin{equation}\label{L_i}
    L_i=\left\{
    \begin{array}{rcl}
        \frac{2(y_i-D)s-s^2}{2\sigma^2}&\text{if $\mathcal{H}_1$}\\
        \frac{2(y_i+D)s-s^2}{2\sigma^2}&\text{if $\mathcal{H}_0$}.
    \end{array}\right.
\end{equation}

 Hence, if sensor $i=H$, the PDF of $L_i$ conditioned on hypothesis $\mathcal{H}_h$ is given by
\begin{equation}\label{Li_H}
        f_{L_i}(l_i|\mathcal{H}_h,i\!=\!H)\!=\! \frac{1}{\sqrt{2\pi\sigma_{h}^2}}\exp\left(-\frac{(L_i-\mu_{h})^2}{\sigma_{h}^2}\right),
\end{equation}
for $h=\{0,1\}$, where $\mu_{1}=\frac{s^2}{2\sigma^2}$, $\mu_{0}=\frac{-s^2}{2\sigma^2}$, $\sigma_{1}^2=\sigma_{0}^2=\frac{s^2}{\sigma^2}=\beta$. Furthermore, if sensor $i=B$, the PDF of $L_i$ conditioned on hypothesis $\mathcal{H}_h$ is given by
\begin{equation}\label{Li_B}
        f_{L_i}(l_i|\mathcal{H}_h,i=B)= \frac{1}{\sqrt{2\pi\nu_{h}^2}}\exp\left(-\frac{(L_i-\eta_{h})^2}{\nu_{h}^2}\right),
\end{equation}
for $h=\{0,1\}$, where $\eta_{0}=\frac{s^2-2Ds}{2\sigma^2}$, $\eta_{1}=\frac{2Ds-s^2}{2\sigma^2}$, $\nu_{0}^2=\nu_{1}^2=\frac{s^2}{\sigma^2}=\beta$.

Therefore, utilizing \eqref{Li_H} and \eqref{Li_B}, the PDF of $L_i$ given $\mathcal{H}_1$ and the PDF of $L_i$ given $\mathcal{H}_0$ are expressed as
\begin{align}\label{eq:f_Li}
    f_{L_i}(l_i|\mathcal{H}_h)&=\alpha_0 f_{L_i}(l_i|\mathcal{H}_h,i=B)\!+\!(1-\alpha_0)f_{L_i}(l_i|\mathcal{H}_h,i=H)\notag\\
    &=\alpha_0 \mathcal{N}(\eta_{h},\nu_{h}^2)\!+\!(1-\alpha_0)\mathcal{N}(\mu_{h},\sigma_{h}^2),
\end{align}
for $h=\{0,1\}$. Here, $\alpha_0$ denotes the probability of a node being Byzantine.
% , where $\phi(x|\mu,\sigma^2)$ is the PDF of $x\sim N(\mu,\sigma^2)$, for simplicity, denoted as $\phi(\mu,\sigma^2)$. 
Let $Z=\sum_{i=1}^NL_{[i]}$ denote the global test statistic. Let $f(Z|H_h)$ denote the Gaussian mixture with PDF given by
\begin{equation}
% \begin{split}
    f(Z|H_h)=\sum_{A_i\in J}(1-\alpha_0)^{N-m}\alpha_0^{m}\mathcal{N}((\mu_h)_{A_t},N\beta),
% \end{split}
\end{equation}
for $h=\{0,1\}$, where $(\mu_h)_{A_i}=\mu_{h}|A_i|+ \eta_{h}|A_i^c|$. Let $J=\{A_1,\dots A_i,\dots ,A_{2^N}\}$ denotes the power set that contains all possible subsets of set $\{1,\dots,N\}$ and $A_i$ is the $i^{th}$ subset of the combination of honest nodes. Also, $|A_i|$ and $|A_i^c|$ are the cardinalities of sets $A_i$ and $A_i^c$, respectively.

Therefore, the detection performance can be evaluated using the probability of detection $P_d^{fc}$ and the probability of false alarm $P_f^{fc}$ given below as
\begin{equation}\label{eq:pd}
    P_d^{fc}=\sum_{A_i\in J}(1-\alpha_0)^{N-m}\alpha_0^{m}Q\left(\frac{\lambda_f-(\mu_1)_{A_i}}{\sqrt{N\beta}}\right)
\end{equation}
and
\begin{equation}\label{eq:pf}
    P_f^{fc}=\sum_{A_i\in J}(1-\alpha_0)^{N-m}\alpha_0^{m}Q\left(\frac{\lambda_f-(\mu_0)_{A_i}}{\sqrt{N\beta}}\right).
\end{equation}
\subsection{Average Number of Transmissions saved for OT-based system under Attack}
We consider the effect of Byzantine attacks on the number of transmissions saved by OT scheme. When the system is under attack, we derive an expression for the average number of transmissions $\bar{N_t}$ in the following theorem. Let $k^*$ denote the minimum number of transmissions needed to make a decision on which  hypothesis is true.

\paragraph*{Theorem 1} The average number of transmissions $\bar{N_t}$ is given as
\begin{equation}\label{eq:expected_N_s}
    \bar{N_t}=\sum_{k=1}^N\pi_1Pr(k^*\geq k|\mathcal{H}_1)+\pi_0Pr(k^*\geq k|\mathcal{H}_0)
\end{equation}
where 
\begin{multline}
\label{eq:prob_exp}
Pr(k^*\geq k|\mathcal{H}_h)\\=E_{\mathbf{L}_{k-1}}\left[F_{|L_i|}(|L_{k-1}||H_h)^{N-k+1}\mathbf{1}_{\{\mathcal{J}\}}\frac{N!}{(N-k+1)!}\right],    
\end{multline}
for $h=\{0,1\}$. Let $F_{|L_i|}(|l_i||H_h)$ be the cumulative distribution dunction (CDF) of $|L_i|$ for $h=\{0,1\}$ provided as
\begin{align}
    &F_{|L_i|}(|l_i||H_h)\notag\\
    =&\alpha_0*\left(Q\left(\frac{-|l_i|-\eta_{h}}{\sqrt{\nu_{h}}}\right)-Q\left(\frac{|l_i|-\eta_{h}}{\sqrt{\nu_{h}}}\right)\right)\nonumber \\
    &+(1-\alpha_0)*\left(Q\left(\frac{-|l_i|-\mu_{h}}{\sqrt{\sigma_{h}}}\right)-Q\left(\frac{|l_i|-\mu_{h}}{\sqrt{\sigma_{h}}}\right)\right),
\end{align}
where $Q(.)$ is Q function. The indicator function $\mathbf{1}_{\{\mathcal{J}\}}$ is 1 when $\mathbf{L}_{k-1}=\{L_1,L_2\dots,L_{k-1}\}$ is in the region $\mathcal{J}$, and 0 otherwise. Here, $\mathcal{J}$ is a hyperplane with $k-1$ dimensions formed by the intersection of three hyperplanes, $\mathcal{J}=\mathcal{L}\bigcap\mathcal{U}\bigcap\mathcal{D}$, given below
\begin{subequations}\label{eq:range}
\begin{align}
        \mathcal{L}&=\left\{ \mathbf{L}_{k-1}:\sum^{k-1}_{i=1}L_{[i]}\leq \lambda+(N-k+1)|L_{[k-1]}|\right\}\\
        \mathcal{U}&=\left\{ \mathbf{L}_{k-1}:\sum^{k-1}_{i=1}L_{[i]}\geq \lambda-(N-k+1)|L_{[k-1]}|\right\}\\
        \mathcal{D}&=\left\{\mathbf{L}_{k-1}:L_1>L_2>\dots>L_{k-1}\right\}
\end{align}
\end{subequations}
\begin{IEEEproof}
 Please see Appendix \ref{Proof_Theo1}. 
\end{IEEEproof}
For a given $k$, we evaluate~\eqref{eq:prob_exp} numerically using the Monte Carlo approach as the following. We generate $M$ i.i.d. realizations of $L_1,L_2,\dots,L_{k-1}$ where the PDF of $L_i$ is given in \eqref{eq:f_Li}, for $\forall{i}\in\{1,2,\dots,k-1\}$. From our experiments, we observe that when $N$ increases, the number of samples $M$ needed to get an accurate evaluation of \eqref{eq:prob_exp} significantly increases. 

Next, we derive the upper bound and the lower bound for the number of transmissions saved by utilizing the OT scheme in the following Theorem. Let $\bar{N_s}^U$ and $\bar{N_s}^L$ denote the upper bound and the lower bound of transmissions saved. %Then we have the following Theorem.
\paragraph*{Theorem 2} When $N$ is sufficiently large, the average number of transmissions saved $\bar{N_s}$ can be bounded as $\bar{N_s}^L\leq\bar{N_s}\leq \bar{N_s}^U$ where
\begin{align}\label{eq:N_s^U}
    &\bar{N_s}^U\notag\\
    =&\sum_{k=1}^{N-1}\sum_{h=0}^1Pr\left(|L_{[k-1]}|\leq\frac{g_U-\lambda}{N-k}|\mathcal{H}_h\right)\notag\\
    &+Pr\left(|L_{[k-1]}|\leq\frac{\lambda-g_L}{N-k}|\mathcal{H}_h\right)\notag\\
    &-Pr\left(|L_{[k-1]}|\leq \min\left(\frac{g_U-\lambda}{N-k},\frac{\lambda-g_L}{N-k}\right)|\mathcal{H}_h\right),
\end{align}
\begin{align}\label{eq:N_s^L}
    &\bar{N_s}^L\notag\\
    =&\sum_{k=1}^{N-1}\sum_{h=0}^1Pr\left(|L_{[k]}|<\frac{g_L-\lambda}{(N-k)}|\mathcal{H}_h\right)\notag\\
    &+Pr\left(|L_{[k]}|<\frac{\lambda-g_U}{(N-k)}|\mathcal{H}_h\right),
\end{align}
and
\begin{equation}
    Pr\left(|L_{[k]}|<W|\mathcal{H}_h\right)=\int_{0}^{W}f_{|L_{[k]}|}(|l_{[k]}||\mathcal{H}_h)\mathrm{d}|l_{[k]}|,
\end{equation}
for $ W\in\{\frac{g_U-\lambda}{N-k},\frac{\lambda-g_L}{N-k},\min(\frac{g_U-\lambda}{N-k},\frac{\lambda-g_L}{N-k}), \frac{g_L-\lambda}{N-k}, \frac{\lambda-g_U}{N-k}\}$.
\begin{IEEEproof}
 Please see Appendix \ref{Proof_Theo2}. 
\end{IEEEproof}
Next, we discuss the optimal attacking strategy for Byzantine nodes and the effect of Byzantine nodes that utilizes the optimal attacking strategy on the OT-based system.
\subsection{Optimal attacking strategy}
From \eqref{eq:pd} and \eqref{eq:pf}, we evaluate the performance of the system utilizing the probability of error, $P_e=\pi_1(1-P_d^{fc})+\pi_0P_f^{fc}$. However, $|\mathcal{J}|$ grows exponentially as $N$ increases. Therefore, it is intractable to evaluate the system performance using $P_e$. Hence, we utilize the deflection coefficient (DC)~\cite{van2004detection} as a surrogate to analyze the best attacking strategy. By minimizing DC, $P_e$ is maximized.

For the system without ordering, denote $\tilde Z=\sum^{N}_{i=1}L_{i}$ as the global statistic. The DC is defined as 
\begin{align}\label{dcjammer}
    D(\tilde Z)=\frac{(\mathbb{E}(\tilde Z|H_1)-\mathbb{E}(\tilde Z|H_0))^2}{Var(\tilde Z|H_0)},
\end{align}where 
\begin{align}
    &\mathbb{E}(\tilde Z|H_1)=\sum\limits_{i=1}^{N}\mathbb{E}(L_i|H_1)\notag\\
    &=\alpha_0\sum\limits_{i=1}^{N}\mathbb{E}(L_i|H_1,i=B)+(1-\alpha_0)\sum\limits_{i=1}^{N}\mathbb{E}(L_i|H_1,i=H)\notag\\
    &=N\frac{s^2-2Ds\alpha_0}{2\sigma^2}\\
    &\mathbb{E}(\tilde Z|H_0)=\sum\limits_{i=1}^{N}\mathbb{E}(L_i|H_0)\notag\\
    &=\alpha_0\sum\limits_{i=1}^{N}\mathbb{E}(L_i|H_0,i=B)+(1-\alpha_0)\sum\limits_{i=1}^{N}\mathbb{E}(L_i|H_0,i=H)\notag\\
    &=N\frac{2Ds\alpha_0-s^2}{2\sigma^2}
\end{align}

From Lemma 1 and above discussion, to maximize the probability of error with ordering, we could minimize the DC without ordering. The value of $D$ given a specific $\alpha_0$ (or the value of $\alpha_0$ given a specific $D$) which minimizes DC is the optimal attacking strength $D^*$ (or the optimal fraction of Byzantine nodes $\alpha_0^*$ in the network). Since the DC is always non-negative, Byzantine nodes want to make $D(\tilde Z)=0$. From \eqref{dcjammer}, when $\mathbb{E}(\tilde Z|H_1)=\mathbb{E}(\tilde Z|H_0)$, we have $D(\tilde Z)=0$. Hence, for a given $\alpha_0$, the optimal attacking strength $D^*$ is given by
\begin{equation}\label{eq:opt_d}
    D^*=\frac{s}{2\alpha_0},
\end{equation}
which is the minimum attacking strength to blind the FC.

\section{Simulation Results}
\label{sec:simulation}
In this section, we present the numerical results. We assume that identical sensors are utilized in the network. Hence, we have $P_{d}=0.9$, $P_{f}=0.1$ for $i=\{1,\dots,N\}$.

\begin{figure}[htbp]
\centerline{\includegraphics[width=\linewidth,height=15em]{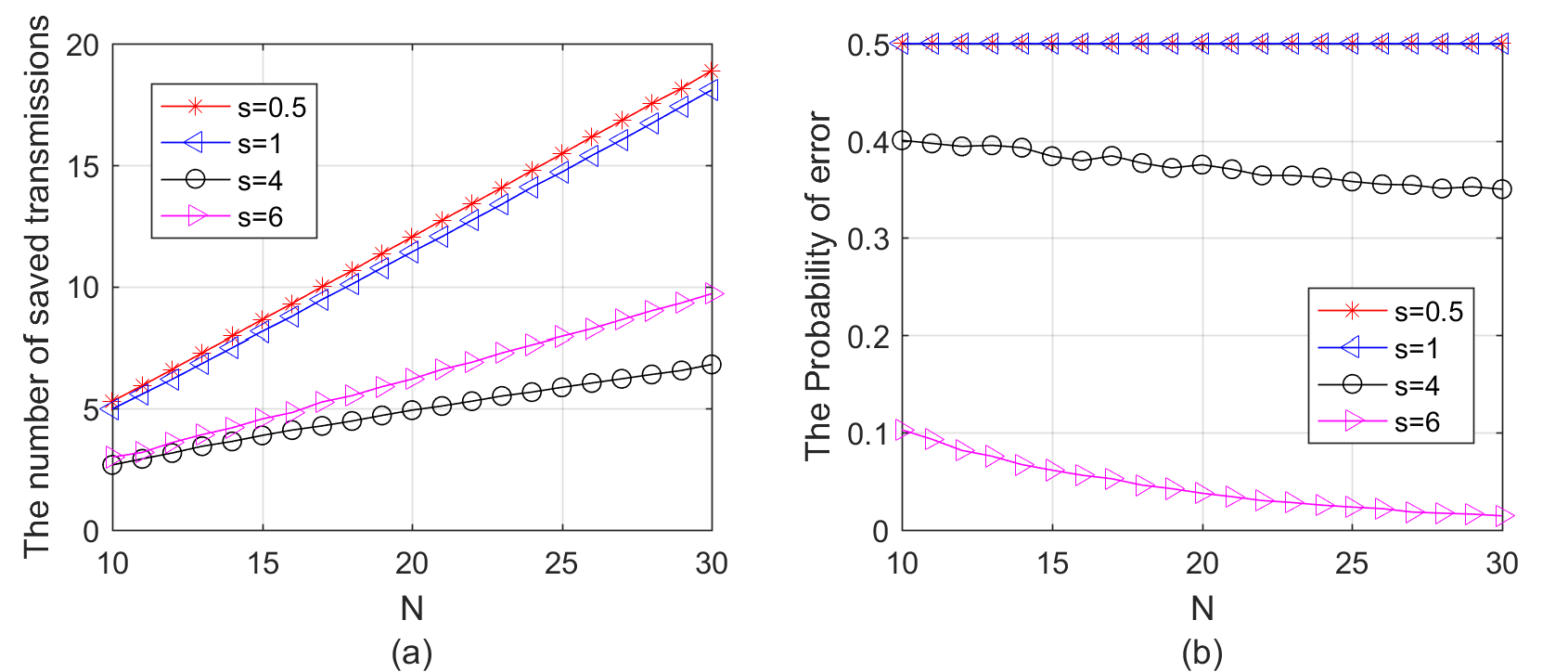}}
\caption{(a) Number of transmissions saved as a function of $N$ for different values of $s$ when $\pi_1=\pi_0=0.5$, $\alpha_0=0.3$ and $D=6$. (b) $P_e$ as a function of $N$ for different values of $s$ when $\pi_1=\pi_0=0.5$, $\alpha_0=0.3$ and $D=6$}
\label{fig:change_N}
\end{figure}

Fig. \ref{fig:change_N}(a) plots the number of transmissions saved as a function of the total number of sensors $N$ for different values of the signal strength $s$ when $\alpha_0=0.3$ and $D=6$. It shows that the number of transmissions saved increases with a decreased $s$ when $s$ is small and increases with an increased $s$ when $s$ is large. Fig.\ref{fig:change_N} (b) shows the probability of error. We observe that the probability of error increases when the signal strength $s$ is small. Although the number of transmissions saved is larger when $s=0.5$ compared to that when $s=4$, the detection performance significantly degrades. This happens because a smaller $s$ leads to a shorter decision interval (or smaller the number of transmissions needed to make the final decision) even with a worse detection performance.

\begin{figure}[htbp]
\centerline{\includegraphics[width=\linewidth,height=15em]{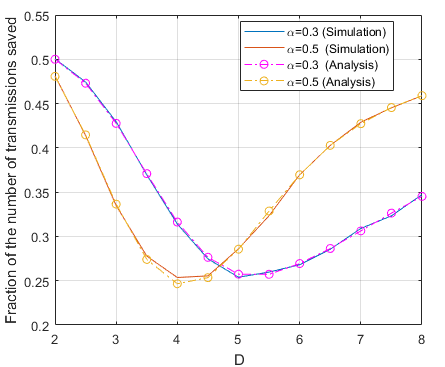}}
\caption{Comparison of $\bar{N_s}/N$ as a function of $D$ for different values of $\alpha_0$ when $\pi_1=\pi_0=0.5$ and $N=10$.}
\label{fig:simulation_analytical}
\end{figure}

Fig. \ref{fig:simulation_analytical} plots the average percentage of saving $\bar{N_s}/N$ as a function of $D$ for different values of $\alpha_0$. 
% It shows that the average percentage of saving first decreases, then increases. 
Initially, $\bar{N_s}/N$ decreases when $D$ increases. However, when $D$ further increases, the FC starts to make wrong decisions and the number of transmissions needed to make the final decision starts to decrease. We also plot the comparison between the results obtained via simulation using Monte Carlo method and analysis using \eqref{eq:expected_N_s}, and observe a good match. 
% We can observe a good match between the results obtained via Monte Carlo method and the one obtained via analytical expression in \eqref{eq:expected_N_s}. 
We also observe that the optimal attacking strength $D^*$ from \eqref{eq:opt_d} is near the critical operation point which minimizes the average percentage of saving. Therefore, the optimal attacking strength $D^*$ from \eqref{eq:opt_d} can not only blind the FC, but also lead to a smaller average percentage of saving. %We also founded that $M=10^8$ samples are needed to give a good match between the simulation results and analytical result for $N=10$. When we want get the good match with a larger $N$, the number of sensors needed significantly increases.
\begin{figure}[htbp]
\centerline{\includegraphics[width=\linewidth,height=15em]{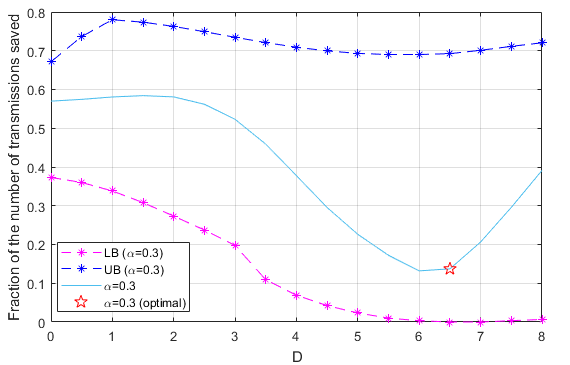}}
\caption{Benchmarking upper and lower bounds for $N_t/N$ as a function of $D$ for different values of $\alpha_0$ with different approaches when $\pi_1=\pi_0=0.5$ and $N=300$.}
\label{fig:UBLB}
\end{figure}

\begin{figure}[htbp]
\centerline{\includegraphics[width=\linewidth,height=15em]{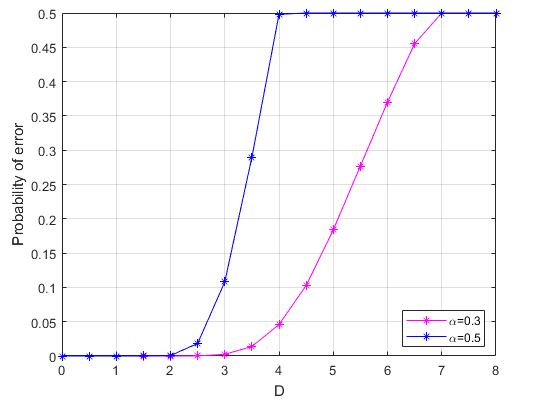}}
\caption{$P_e$ as a function of $D$ with different values of $\alpha_0$ when $\pi_1=\pi_0=0.5$ and $N=300$.}
\label{fig:pe}
\end{figure}

Fig. \ref{fig:UBLB} shows the UB and LB for the average percentage of saving as a function of the attacking strength $D$ when $N=300$. We observe that the LB obtained numerically is very closed to the one obtained in
\eqref{eq:N_s^L}. Although we have a relatively larger gap between the UB obtained numerically and the one obtained in \eqref{eq:N_s^U}, it shows a similar trend as that of the average percentage of saving. We also observe that the optimal attacking strength $D^*$ from \eqref{eq:opt_d} is closer to the critical operation point that minimizes $\bar{N_s}$. Fig. \ref{fig:pe} plots probability of error. We observe that a larger $D$ is needed to blind the FC when $\alpha_0$ decreases.

\section{Conclusion}
\label{sec:conclusion}
In this paper, we investigated the effect of Byzantine attacks on the performance of the system using OT scheme for binary hypothesis testing problem. We derived the probability of error for the OT-based system under attacks and the analytical number of transmissions saved. We also obtained the upper bound and lower bound of the number of transmission saved for the system under attacks. The optimal attacking strategy is investigated and the performance of the system is evaluated based on the deflection coefficient. The simulation results showed that the Byzantine nodes can both maximize the probability of error and significantly increase the number of transmissions needed to make the final decision when they adopt the optimal attacking strategy.
\appendices
\section{Proof of Theorem 1}\label{Proof_Theo1}
Let $\bar{N_t}$ denote the average number of transmissions in the network. $\bar{N_t}$ is given as
\begin{subequations}
\begin{align}
    \bar{N_t}&=E(k^*)\\
    &=\sum^N_{k=1}kPr(k^*=k)\\
    &=\sum_{k=1}^NPr(k^*\geq k)\\
    &=\sum_{k=1}^NPr(k^*\geq k|\mathcal{H}_0)\pi_0+Pr(k^*\geq k|\mathcal{H}_1)\pi_1
\end{align}
\end{subequations}
where $Pr(k^*\geq k)$ is the probability that at least $k$ transmissions in the network are needed to make the final decision. Next Lemma helps us to obtain the probability of the event that at least $k$ transmissions are required to make the final decision.
\begin{lemma} The FC can not decided $\mathcal{H}_1$ or $\mathcal{H}_0$ until the FC has received at least $k$ transmissions  if
\begin{subequations}
\begin{align}
        \sum^{k-1}_{i=1}L_{[i]}&\leq \lambda+(N-k+1)|L_{[k-1]}|\\
        \sum^{k-1}_{i=1}L_{[i]}&\geq \lambda-(N-k+1)|L_{[k-1]}|
\end{align}
\end{subequations}
\end{lemma}
\begin{IEEEproof}
 When the FC received the first $(k-1)$ LLRs, i.e, $[L_{[1]},L_{[2]},\dots,L_{[k-1]}]$, we discuss the cases that the FC can not decide $\mathcal{H}_1$ and the FC can not decide $\mathcal{H}_0$, respectively. 
 \underline{Can not Decide $\mathcal{H}_0$:}

Recall that $|L_{[1]}|\geq|L_{[2]}|\dots\geq{|L_{[N]}|}$, we have
\begin{equation}
    Z\leq\underbrace{\sum^{k-1}_{i=1}L_{[i]}+(N-k+1)|L_{[k-1]}|}_{\eta_U}
\end{equation}
Obviously, the FC is not able to decide $\mathcal{H}_0$ when $\eta_U>\lambda$. Moreover, \eqref{eq:U_for_k-1} shows that if the FC doesn't decide $\mathcal{H}_0$ after receiving the first $(k-1)$ LLRs, it can't decide $\mathcal{H}_0$ after receiving the first $(k-2)$ observations.
\begin{subequations}\label{eq:U_for_k-1}
\begin{align}
    \eta_U=&\sum^{k-1}_{i=1}L_{[i]}+(N-k+1)|L_{[k-1]}|\\
    =&\sum^{k-2}_{i=1}L_{[i]}+L_{[k-1]}+(N-k+2)|L_{[k-2]}|\notag\\
    &+(N-k+1)|L_{[k-1]}|-(N-k+2)|L_{[k-2]}|\\
    =&\sum^{k-2}_{i=1}L_{[i]}+(N-k+2)|L_{[k-2]}|+(N-k+1)\notag\\
    &*(|L_{[k-1]}|-|L_{[k-2]}|)+(L_{[k-1]}-|L_{[k-2]}|)\label{eq:U_for_k-1_1}\\
    \overset{imply}{\Longrightarrow}&\sum^{k-2}_{i=1}L_{[i]}+(N-k+2)|L_{[k-2]}|>\lambda\label{eq:U_for_k-1_2}
\end{align}
\end{subequations}
As $|L_{[k-1]}|\leq|L_{[k-2]}|$ and $L_{[k-1]}\leq|L_{[k-1]}|\leq|L_{[k-2]}|$, we have $|L_{[k-1]}|-|L_{[k-2]}|\leq 0$ and $L_{[k-1]}-|L_{[k-2]}|\leq 0$ in \eqref{eq:U_for_k-1_2}. Hence, we can obtain that \eqref{eq:U_for_k-1_1}$\geq\lambda$ implies \eqref{eq:U_for_k-1_2}$\geq\lambda$. Following the similar procedure as shown in \eqref{eq:U_for_k-1}, we are able to conclude that if the FC can't decide $\mathcal{H}_0$ after receiving the first $(k-1)$ LLRs, it can't decide $\mathcal{H}_0$ after receiving 0 or 1 or $\dots$, or $(k-2)$ observations.

\underline{Can not Decide $\mathcal{H}_1$:}

we have the following inequality in \eqref{eq:inequ2} after the FC has received the first $(k-1)$ LLRs. Obviously, the FC can not decide $\mathcal{H}_1$ when $\eta_L<\lambda$.
\begin{equation}\label{eq:inequ2}
    \underbrace{\sum^{k-1}_{i=1}L_{[i]}-(N-k+1)|L_{[k-1]}|}_{\eta_L}\leq Z
\end{equation}
Following the similar procedure as shown in \eqref{eq:U_for_k-1}, we can prove that if the FC can't decide $\mathcal{H}_1$ after receiving the first $(k-1)$ largest LLRs, it can't decide $\mathcal{H}_1$ after receiving 0 or 1 or $\dots$, or $(k-2)$ observations. The proof for this is similar as above and is skipped. 
\end{IEEEproof} 

To evaluate $Pr(k^*\geq k|\mathcal{H}_h)$, we have
\begin{align}\label{eq:int_f}
    &Pr(k^*\geq k|\mathcal{H}_h)\notag\\
    &=\int_{\mathbf{l}_{k-1}\in\mathcal{J}}f_{\mathbf{L_{[k-1]}}}(l_{[1]},\dots,l_{[k-1]}|\mathcal{H}_h)\mathrm{d}l_{1}\dots\mathrm{d}l_{k-1},
\end{align}
where $f_{\mathbf{L_{[k-1]}}}(l_{[1]},\dots,l_{[k-1]}|\mathcal{H}_h)$ is the joint pdf of $l_{[1]},l_{[2]},\dots,l_{[k-1]}$ given $\mathcal{H}_h$ for $h=0,1$. According to \cite{david2004order}, the joint pdf of $l_{[1]},l_{[2]},\dots,l_{[k-1]}$ given $\mathcal{H}_h$ is given as
\begin{align}\label{eq:f}
    &f_{\mathbf{L_{[k-1]}}}(l_{[1]},\dots,l_{[k-1]}|\mathcal{H}_h)=\frac{N!}{(N-k+1)!}\notag\\
    &*\left[\prod_{i=1}^{k-1}f_{L_{i}}(l_{i}|\mathcal{H}_h)\right]\left[F_{|L_i|}(|l_i||H_h)\right]^{N-k+1}\mathbf{1}_{\{\mathcal{J}\}}
\end{align}
where $\mathcal{J}=\mathcal{L}\bigcap\mathcal{U}\bigcap\mathcal{D}$ is the intersection of hyperplanes $\mathcal{L}$, $\mathcal{U}$ and $\mathcal{D}$ and $F_{|L_i|}(|l_i||H_h)$ is the cdf of $|L_i|$ for $h=0,1$. By substituting \eqref{eq:f} in \eqref{eq:int_f} and utilizing the law of total expectation, \eqref{eq:int_f} can be rewritten as
\begin{align}
    Pr&(k^*\geq k|\mathcal{H}_h)\notag\\
    &=E_{\mathbf{L}_{[k-1]}}\left[\frac{N!}{(N-k+1)!}\left[F_{|L_i|}(|l_i||H_h))\right]^{N-k+1}\mathbf{1}_{\{\mathcal{J}\}}\right]
\end{align}
for $h=0,1$.

% \section{Appendix}
\section{Proof of Lemma~\ref{lem:det_perf_ot}}
\label{sec:proof_det_perf_ot}
 Let $Z_U$, $Z_L$ denote the upper bound and lower bound of $Z=\sum_{i=1}^NL_{[i]}$, respectively. Due to the fact that $|L_{[1]}|>|L_{[2]}|>\dots>|L_{[N]}|$, we have 
\begin{equation}\label{eq:Z_U}
    Z_U=\sum_{i=1}^{k_U^*}L_{[i]}+(N-k_U^*)|L_{[k_U^*]}|\geq \sum_{i=1}^NL_{[i]}
\end{equation}
and
\begin{equation}\label{eq:Z_l}
    Z_L=\sum_{i=1}^{k_L^*}L_{[i]}-(N-k_L^*)|L_{[k_L^*]}|\leq \sum_{i=1}^NL_{[i]},
\end{equation}
where $k_U^*=\min\limits_{1\leq k\leq N}\left\{\sum_{i=1}^kL_{[i]}\leq \lambda -(N-k)|L_{[k]}|\right\}$ and $k_L^*=\min\limits_{1\leq k\leq N}\left\{\sum_{i=1}^kL_{[i]}\geq \lambda -(N-k)|L_{[k]}|\right\}$. If $Z_U<\lambda$, the FC can decide hypothesis $\mathcal{H}_0$, and if $Z_L>\lambda$, the FC can decide hypothesis $\mathcal{H}_1$. Intuitively, we can conclude that $\mathbb{P}(Z_L>\lambda|Z>\lambda,\mathcal{H}_j)=1$, since at least $k_L^*=N$ guarantees this equality is true. We also have $\mathbb{P}(Z>\lambda|Z_L>\lambda,\mathcal{H}_j)=1$. 

Hence, we can calculate $\mathbb{P}(Z_L>\lambda|\mathcal{H}_j)$ according to Bayesian rule given as
% \begin{equation}
\begin{align}
        Pr(Z_L>\lambda|\mathcal{H}_j)&=\frac{Pr(Z_L>\lambda|Z>\lambda,\mathcal{H}_j)Pr(Z>\lambda|\mathcal{H}_j)}{Pr(Z>\lambda|Z_L>\lambda,\mathcal{H}_j)} \nonumber \\
    &=Pr(Z>\lambda|\mathcal{H}_j).
\end{align}
% \end{equation}
Similarly, we obtain $Pr(Z_U<\lambda|\mathcal{H}_j)=Pr(Z<\lambda|\mathcal{H}_j)$. Hence, the probability of error of the OT-based system is given as
% \begin{subequations}
\begin{align}
    P_e^{(OT)}&=\pi_0Pr(Z_L>\lambda|\mathcal{H}_0)+\pi_1Pr(Z_U<\lambda|\mathcal{H}_1) \nonumber \\
    &=\pi_0Pr(Z>\lambda|\mathcal{H}_0)+\pi_1Pr(Z<\lambda|\mathcal{H}_1)=P_e^{(opt)},
    \end{align}
% \end{subequations}
where $P_e^{(opt)}$ is the probability of error of the system without OT scheme.

\section{Proof of Theorem 2}\label{Proof_Theo2}
% \begin{IEEEproof}
Let $\bar{N_s}$ denote the average number of transmissions saved in the network given as
\begin{subequations}\label{eq:expected_N_s}
\begin{align}
    \bar{N_s}&=E(k^*)\\
    &=\sum^N_{k=1}(N-k)Pr(k^*=k)\\
    &=\sum_{k=1}^{N-1}Pr(k^*\leq k)\\
    &=\sum_{k=1}^{N-1}Pr(k^*\leq k|\mathcal{H}_0)\pi_0+Pr(k^*\leq k|\mathcal{H}_1)\pi_1.
\end{align}
\end{subequations}
Next, we use the following lemma from \cite[Chapter 5]{david2004order} to prove Theorem 2.

\begin{lemma}
\label{lem:cauchy}
According to Cauchy's inequality, we have
\begin{equation}
    |\sum c_i(L_{[i]}-\bar{L})|\leq[\sum(c_i-\bar{c})^2(N-1)v]^{\frac{1}{2}}
\end{equation}
in terms of empirical mean $\bar{L}$ and empirical variance $v$ for any constants $\{c_i\}_{i=1}^N$. If $c_i$ is non-increasing when $i$ increases, the bound is sharp.
\end{lemma}
% \paragraph*{Lemma 2}

From Lemma \ref{lem:cauchy}, we have
\begin{equation}\label{eq:UBLB}
    |\sum_{i=1}^k L_{[i]}-k\bar{L}|\leq[\sum(c_i-\bar{c})^2(N-1)v]^{\frac{1}{2}}
\end{equation}
if we let $c_1=c_2=\dots=c_k=1$ and $c_{k+1}=\dots=c_N=0$. Hence, from \eqref{eq:UBLB}, the lower bound and the upper bound of $\sum_{i=1}^kL_{[i]}$ are given by
\begin{equation}
    g_L\leq\sum_{i=1}^kL_{[i]}\leq g_U,
\end{equation}
where $g_L=-[\sum(c_i-\bar{c})^2(N-1)v]^{\frac{1}{2}}+k\bar{L}$ and $g_U=[\sum(c_i-\bar{c})^2(N-1)v]^{\frac{1}{2}}+k\bar{L}$. 

\underline{LB of $\bar{N_s}$:} When the FC decides $\mathcal{H}_1$ in at most $k$ transmissions given hypothesis $\mathcal{H}_h$, we have
\begin{equation}\label{eq:at_most_k}
    Pr(k^*\leq k|\mathcal{H}_h)=Pr(\sum_{i=1}^kL_{[k]}> \lambda+(N-k)|L_{[k]}||\mathcal{H}_h).
\end{equation}
for $h=0,1$. It is easy to show that $g_L>\lambda+(N-k)|L_{[k]}|$ implies $\sum_{i=1}^kL_{[k]}> \lambda+(N-k)|L_{[k]}|$. Hence, from \eqref{eq:at_most_k}, we get
\begin{equation}
    Pr(k^*\leq k|\mathcal{H}_h)\geq Pr(g_L>\lambda+(N-k)|L_{[k]}||\mathcal{H}_h)
\end{equation}

Similarly, when the FC decides $\mathcal{H}_0$ in at most $k$ transmissions given hypothesis $\mathcal{H}_h$, we get
\begin{equation}\label{eq:g_U_LB}
    Pr(k^*\leq k|\mathcal{H}_h)\geq Pr(g_U<\lambda-(N-k)|L_{[k]}||\mathcal{H}_h)
\end{equation}
The inequality in \eqref{eq:g_U_LB} is true due to the fact that $g_U<\lambda-(N-k)|L_{[k]}|$ implies $\sum_{i=1}^kL_{[i]}<\lambda-(N-k)|L_{[k]}|$. Substituting $Pr(k^*\leq k|\mathcal{H}_0)$ and $Pr(k^*\leq k|\mathcal{H}_1)$ in \eqref{eq:expected_N_s} with their lower bounds $Pr(g_L>\lambda+(N-k)|L_{[k]}||\mathcal{H}_h)$ and $Pr(g_U<\lambda-(N-k)|L_{[k]}||\mathcal{H}_h)$, respectively, we get
\begin{align}\label{eq:N_s_LB_H1_H0}
    \bar{N_s}\geq&\sum_{k=1}^{N-1}\sum_{h=0}^1Pr(g_L>\lambda+(N-k)|L_{[k]}||\mathcal{H}_h)\notag\\
    &+Pr(g_U<\lambda-(N-k)|L_{[k]}||\mathcal{H}_h)
\end{align}
A Monte Carlo approach can be utilized to evaluate $Pr(g_L>\lambda+(N-k)|L_{[k]}||\mathcal{H}_h)$ and $Pr(g_U>\lambda-(N-k)|L_{[k]}||\mathcal{H}_h)$. We generate $M_2$ realizations of $L_{[1]},L_{[2]},\dots,L_{[N]}$ so that the empirical mean $\bar{L}$ and the empirical variance $v$ can be calculated. When $N$ is sufficiently large, sample mean $\bar{L}$ approaches the population mean. The population mean under $\mathcal{H}_1$ and the one under $\mathcal{H}_0$ are expressed, respectively, as
\begin{subequations}
\begin{align}
    \delta_1&=E[L_i|\mathcal{H}_1]=\alpha_0\eta_{1}+(1-\alpha_0)\mu_{1}\\
    \delta_0&=E[L_i|\mathcal{H}_0]=\alpha_0\eta_{0}+(1-\alpha_0)\mu_{0}
\end{align}
\end{subequations}
The population variance under $\mathcal{H}_h$ is expressed as
\begin{subequations}
\begin{align}
    \zeta_h^2=&E[L_i^2|\mathcal{H}_h]-\delta_h^2\notag\\
    =&E[L_i^2|\mathcal{H}_h]-(\alpha_0\eta_{h}+(1-\alpha_0)\mu_{h})^2,
\end{align}
\end{subequations}
where
\begin{subequations}
\begin{align}
    E[L_i^2|\mathcal{H}_h]=&\alpha_0 E[L_i^2|\mathcal{H}_h,i=B]+(1-\alpha_0)E[L_i^2|\mathcal{H}_h,i=H]\notag\\
    =&\beta+\alpha_0\eta_{h}^2+(1-\alpha_0)\mu_{h}^2
\end{align}
\end{subequations}
for $h=0,1$. Substituting the parameters $(\bar{L},v)$ in \eqref{eq:N_s_LB_H1_H0} with parameters $(\delta_h,\frac{N}{N-1}\zeta_h^2)$ under $\mathcal{H}_h$ for $h=0,1$ yields
\begin{align}\label{eq:N_s_LB_H1_H0_2}
    \bar{N_s}\geq&\sum_{k=1}^{N-1}\sum_{h=0}^1Pr\left(|L_{[k]}|<\frac{g_L-\lambda}{(N-k)}|\mathcal{H}_h\right)\notag\\
    &+Pr\left(|L_{[k]}|<\frac{\lambda-g_U}{(N-k)}|\mathcal{H}_h\right).
\end{align}
where
\begin{align}\label{eq:Pr_GL}
    &Pr\left(|L_{[k]}|<\frac{g_L-\lambda}{(N-k)}|\mathcal{H}_h\right)\notag\\
    &=\int_{0}^{\frac{g_L-\lambda}{(N-k)}}f_{|L_{[k]}|}(|l_{[k-1]}||\mathcal{H}_h)\mathrm{d}|l_{[k]}|
\end{align}
and
\begin{align}\label{eq:Pr_GU}
    &Pr\left(|L_{[k]}|<\frac{\lambda-g_U}{(N-k)}|\mathcal{H}_h\right)\notag\\
    &=\int_{0}^{\frac{\lambda-g_U}{(N-k)}}f_{|L_{[k]}|}(|l_{[k-1]}||\mathcal{H}_h)\mathrm{d}|l_{[k]}|.
\end{align}
It is given in closed form as \cite{david2004order}
\begin{align}
    &f_{L_{[k-1]}}(l_{[k-1]}|\mathcal{H}_h)\notag\\
    =&Nf_{L}(l_{[k-1]}|\mathcal{H}_h){N-1\choose k-1}F_{L}(l_{[k-1]}|\mathcal{H}_h)^{(N-k)}\notag\\
    &*(1-F_{L}(l_{[k-1]}|\mathcal{H}_h))^{(k-1)}.
\end{align}
Hence, the pdf of $f_{|L_{[k-1]}|}(l_{[k-1]}|\mathcal{H}_h)$ is given by
\begin{align}\label{eq:f_|L_k-1|}
    &f_{|L_{[k-1]}|}(l_{[k-1]}|\mathcal{H}_h)\notag\\
    =&\frac{\mathrm{d}Pr(|L_{[k-1]}|\leq l_{[k-1]})}{\mathrm{d}l_{[k-1]}}\notag\\
    =&\left\{
    \begin{array}{lcl}
         f_{L_{[k-1]}}(l_{[k-1]}|\mathcal{H}_h)-f_{L_{[k-1]}}(-l_{[k-1]}|\mathcal{H}_h)& \text{if $l_{[k-1]}\geq 0$} \\
         0& \text{if $l_{[k-1]}<0$}
    \end{array}\right.
\end{align}

Substituting \eqref{eq:f_|L_k-1|} in $\eqref{eq:Pr_GL}$ and \eqref{eq:Pr_GU}, we are able to evaluate $Pr(|L_{[k]}|<\frac{g_L-\lambda}{(N-k)}|\mathcal{H}_h)$ and $Pr(|L_{[k]}|<\frac{\lambda-g_U}{(N-k)}|\mathcal{H}_h)$. The lower bound of the number of transmissions saved can be obtained by utilizing \eqref{eq:N_s_LB_H1_H0_2}.
\underline{UB of $\bar{N_s}$:} When the FC decides $\mathcal{H}_1$ in at most $k$ transmissions given hypothesis $\mathcal{H}_h$, it is easy to show that $\sum_{i=1}^kL_{[k]}> \lambda+(N-k)|L_{[k]}|$ implies $g_U>\lambda+(N-k)|L_{[k]}|$. Hence, from \eqref{eq:at_most_k}, we get
\begin{equation}
    Pr(g_U>\lambda+(N-k)|L_{[k]}||\mathcal{H}_h)\geq Pr(k^*\leq k|\mathcal{H}_h)
\end{equation}

Similarly, when the FC decides $\mathcal{H}_0$ in at most $k$ transmissions given hypothesis $\mathcal{H}_h$, we get
\begin{equation}\label{eq:g_U_LB_2}
    Pr(g_L<\lambda-(N-k)|L_{[k]}||\mathcal{H}_h)\geq Pr(k^*\leq k|\mathcal{H}_h).
\end{equation}
The inequality in \eqref{eq:g_U_LB_2} is true due to the fact that $\sum_{i=1}^kL_{[i]}<\lambda-(N-k)|L_{[k]}|$ implies $g_L<\lambda-(N-k)|L_{[k]}|$. Hence, we have
\begin{align}\label{eq:N_s_UB_H1_H0}
    \bar{N_s}\leq&\sum_{k=1}^{N-1}\sum_{h=0}^1Pr(g_U>\lambda+(N-k)|L_{[k]}|\notag\\
    &\text{ or }g_L<\lambda-(N-k)|L_{[k]}||\mathcal{H}_h),
\end{align}
where
\begin{equation}
    \begin{split}
    &Pr(g_U>\lambda+(N-k)|L_{[k]}|\text{ or }g_L<\lambda-(N-k)|L_{[k]}||\mathcal{H}_h)\\
    =&Pr(g_U>\lambda+(N-k)|L_{[k]}||\mathcal{H}_h)+\\
    &Pr(g_L<\lambda-(N-k)|L_{[k]}||\mathcal{H}_h)-\\
    &Pr(g_U>\lambda+(N-k)|L_{[k]}|\text{ and }g_L<\lambda-(N-k)|L_{[k]}||\mathcal{H}_h)\\
    =&Pr\left(|L_{[k-1]}|\leq\frac{g_U-\lambda}{N-k}|\mathcal{H}_h\right)+Pr\left(|L_{[k-1]}|\leq\frac{\lambda-g_L}{N-k}|\mathcal{H}_h\right)\\
    &-Pr\left(|L_{[k-1]}|\leq \min\left(\frac{g_U-\lambda}{N-k},\frac{\lambda-g_L}{N-k}\right)|\mathcal{H}_h\right).
    \end{split}
\end{equation}
Following the similar procedure when we obtain the LB of $\bar{N_s}$, we can get the UB of $\bar{N_s}$ by either Monte Carlo approach or theoretical approach if $N$ is sufficiently large. Then, we can obtain the UB and the LB in Theorem 2.
% \end{IEEEproof}

\bibliography{refer.bib}
\bibliographystyle{IEEEtran}

\end{document}